# Citation analysis may severely underestimate the impact of clinical research as compared to basic research

Nees Jan van Eck[1], Ludo Waltman[1], Anthony F.J. van Raan[1], Robert J.M. Klautz[2], and Wilco C. Peul[3]

[1] Leiden University, Centre for Science and Technology Studies, Leiden, The Netherlands
{ecknjpvan, waltmanlr, vanraan}@cwts.leidenuniv.nl

[2] Leiden University Medical Center, Department of Cardiothoracic Surgery, Leiden, The Netherlands
r.j.m.klautz@lumc.nl

[3] Leiden University Medical Center, Department of Neurosurgery, Leiden, The Netherlands
w.c.peul@lumc.nl

*Background*

Citation analysis has become an important tool for research performance assessment in the medical sciences. However, different areas of medical research may have considerably different citation practices, even within the same medical field. Because of this, it is unclear to what extent citation-based bibliometric indicators allow for valid comparisons between research units active in different areas of medical research.

*Methodology*

A visualization methodology is introduced that reveals differences in citation practices between medical research areas. The methodology extracts terms from the titles and abstracts of a large collection of publications and uses these terms to visualize the structure of a medical field and to indicate how research areas within this field differ from each other in their average citation impact.

*Results*

Visualizations are provided for 32 medical fields, defined based on journal subject categories in the Web of Science database. The analysis focuses on three fields: *Cardiac & cardiovascular systems*, *Clinical neurology*, and *Surgery*. In each of these fields, there turn out to be large differences in citation practices between research areas. Low-impact research areas tend to focus on clinical intervention research, while high-impact research areas are often more oriented on basic and diagnostic research.

*Conclusions*

Popular bibliometric indicators, such as the *h*-index and the impact factor, do not correct for differences in citation practices between medical fields. These indicators therefore cannot be used to make accurate between-field comparisons. More sophisticated bibliometric indicators do correct for field differences but still fail to take into account within-field heterogeneity in citation practices. As a consequence, the citation impact of clinical intervention research may be substantially underestimated in comparison with basic and diagnostic research.

## 1. Introduction

Citation analysis is widely used in the assessment of research performance in the medical sciences (Patel et al., 2011). Especially the *h*-index (Hirsch, 2005) and the impact factor (Chew, Villanueva, & Van der Weyden, 2007; Garfield, 1996, 2006) are extremely popular bibliometric indicators. However, the use of these indicators for performance assessment has important limitations. In particular, both the *h*-index and the impact factor fail to take into account the enormous differences in citation practices between fields of science (e.g., Radicchi, Fortunato, & Castellano, 2008). For instance, the average length of the reference list of a publication is much larger in molecular biology than in mathematics. As a consequence, publications in molecular biology on average are cited much more frequently than publications in mathematics. This difference can be more than an order of magnitude (Waltman, Van Eck, Van Leeuwen, Visser, & Van Raan, 2011a).

More sophisticated bibliometric indicators used by professional bibliometric centers perform a normalization to correct for differences in citation practices between fields of science (e.g., Glänzel, Thijs, Schubert, & Debackere, 2009; Waltman, Van Eck, Van Leeuwen, Visser, & Van Raan, 2011b). These field-normalized indicators typically rely on a field classification system in which the boundaries of fields are explicitly defined (e.g., the journal subject categories in the Web of Science database). Unfortunately, however, practical applications of field-normalized indicators often suggest the existence of differences in citation practices not only between but also within fields of science. As shown in this paper, this phenomenon can be observed especially clearly in medical fields, in which the citation impact of clinical intervention research may be substantially underestimated in comparison with basic and diagnostic research. Within-field heterogeneity in citation practices is not corrected for by field-normalized bibliometric indicators and therefore poses a serious threat to the accuracy of these indicators.

This paper presents an empirical analysis of the above problem, with a focus on the medical sciences. An advanced visualization methodology is used to show how citation practices differ between research areas within a medical field. In particular, substantial differences are revealed between basic and diagnostic research areas on the one hand and clinical intervention research areas on the other hand. Implications of the analysis for the use of bibliometric indicators in the medical sciences are discussed.

## 2. Methodology

The analysis reported in this paper starts from the idea that drawing explicit boundaries between research areas, for instance between basic and clinical areas, is difficult and would require many arbitrary decisions, for instance regarding the treatment of multidisciplinary topics that are in between multiple areas. To avoid the difficulty of drawing explicit boundaries between research areas, the methodology adopted in this paper relies strongly on the use of visualization. The methodology uses so-called term maps (e.g., Van Eck & Waltman, 2011; Waaijer, Van Bochove, & Van Eck, 2010, 2011) to visualize scientific fields. A term map is a two-dimensional representation of a field in which strongly related terms are located close to each other and less strongly related terms are located further away from each other. A term map provides an overview of the structure of a field. Different areas in a map correspond with different subfields or research areas. In the term maps presented in this paper, colors are used to indicate differences in citation practices between research areas. For each term in a map, the color of the term is determined by the average citation impact of the publications in which the term occurs. We note that the use of visualization to analyze the structure and development of scientific fields has a long history (e.g., Börner, 2010), but visualization approaches have not been used before to study differences in citation practices between research areas. The use of term maps, also referred to as co-word maps, has a 30-year history, with early contributions dating back to the 1980s and the beginning of the 1990s (e.g., Peters & Van Raan, 1993; Rip & Courtial, 1984; Tijssen & Van Raan, 1989).

The first methodological step is the definition of scientific fields. This study uses data from the Web of Science (WoS) bibliographic database. This database has a good coverage of the medical literature (Moed, 2005) and is the most popular data source for professional bibliometric analyses. Because of their frequent use in field-normalized bibliometric indicators, the journal subject categories in the WoS database are employed to define fields. There are about 250 subject categories in the WoS database, covering disciplines in the sciences, the social sciences, and the arts and humanities. The analyses reported in this paper are based on all publications in a particular subject category that are classified as *article* or *review* and that were published between 2006 and 2010. For each publication, citations are counted until the end of 2011.

Using natural language processing techniques, the titles and abstracts of the publications in a field are parsed. This yields a list of all noun phrases (i.e., sequences of nouns and adjectives) that occur in these publications. An additional algorithm (Van Eck & Waltman, 2011) selects the 2000 noun phrases that can be regarded as the most characteristic terms of the field. This algorithm aims to filter out general noun phrases, like for instance *result*, *study*, *patient*, and *clinical evidence*. Filtering out these noun phrases is crucial. Due to their general meaning, these noun phrases do not relate specifically to one topic, and they therefore tend to distort the structure of a term map. Apart from excluding general noun phrases, noun phrases that occur only in a small number of publications are excluded as well. This is done in order to obtain sufficiently robust results. The minimum number of publications in which a noun phrase must occur depends on the total number of publications in a field. For the three fields discussed in the next section, thresholds between 70 and 135 publications were used.

Given a selection of 2000 terms that together characterize a field, the next step is to determine the number of publications in which each pair of terms co-occurs. Two terms are said to co-occur in a publication if they both occur at least once in the title or abstract of the publication. The larger the number of publications in which two terms co-occur, the stronger the terms are considered to be related to each other. In neuroscience, for instance, *Alzheimer* and *short-term memory* may be expected to co-occur a lot, indicating a strong relation between these two terms. The matrix of term co-occurrence frequencies serves as input for the VOS mapping technique (Van Eck, Waltman, Dekker, & Van den Berg, 2010). This technique determines for each term a location in a two-dimensional space. Strongly related terms tend to be located close to each other in the two-dimensional space, while terms that do not have a strong relation are located further away from each other. The VOS mapping technique is closely related to the technique of multidimensional scaling (e.g., Borg & Groenen, 2005), but for the purpose of creating term maps the VOS mapping technique has been shown to yield more satisfactory results, as discussed by Van Eck et al. (2010). It is important to note that in the interpretation of a term map only the distances between terms are relevant. A map can be freely rotated, because this does not affect the inter-term distances. This also implies that the horizontal and vertical axes have no special meaning.

In the final step, the color of each term is determined. First, in order to correct for the age of a publication, each publication's number of citations is divided by the average number of citations of all publications that appeared in the same year. This yields a publication's normalized citation score. A score of 1 means that the number of citations of a publication equals the average of all publications that appeared in the same field and in the same year. Next, for each of the 2000 terms, the normalized citation scores of all publications in which the term occurs (in the title or abstract) are averaged. The color of a term is determined based on the resulting average score. Colors range from blue (average score of 0) to green (average score of 1) to red (average score of 2 or higher). Hence, a blue term indicates that the publications in which a term occurs have a low average citation impact, while a red term indicates that the underlying publications have a high average citation impact. The VOSviewer software (Van Eck & Waltman, 2010) is used to visualize the term maps resulting from the above steps.[1]

## 3. Results

Figures 1, 2, and 3 show the term maps obtained for the WoS fields *Cardiac & cardiovascular systems*, *Clinical neurology*, and *Surgery*. These fields were selected because they match well with our areas of expertise. The maps are based on, respectively, 75,314, 105,405, and 141,155 publications from the period 2006–2010. Only a limited level of detail is offered in Figures 1, 2, and 3. To explore the term maps in full detail, the reader is invited to use the interactive versions of the maps that are available on a webpage.[2] The webpage also provides maps of 29 other medical fields as well as of all 32 medical fields taken together.

[1] The VOSviewer software is freely available at www.vosviewer.com.

[2] The interactive maps can be found at www.neesjanvaneck.nl/basic_vs_clinical/.

The term maps shown in Figures 1, 2, and 3 all indicate a clear distinction between different research areas. Clinical research areas tends to be located mainly in the left part of a map and basic research areas mainly in the right part, although making a perfect distinction between basic and clinical research areas is definitely not possible. The basic-clinical distinction is best visible in the *Cardiac & cardiovascular systems* and *Clinical neurology* maps (Figures 1 and 2), in which the left part consists of clinical intervention research areas (e.g., cardiac surgery and neurosurgery) while the right part includes important basic and diagnostic research areas (e.g., cardiology and neurology). The *Surgery* map (Figure 3) gives a somewhat different picture, probably because of the more clinical focus of surgical research. In this map, clinical research areas (e.g., orthopedic surgery, oncological surgery, and cardiac surgery) are concentrated in the left, middle, and upper parts, while research areas with a more basic focus can be found in the lower-right part.

Connections between basic research areas on the one hand and clinical research areas on the other hand are also visible in the term maps. The maps display ‘bridges’ that seem to represent translational research, that is, research aimed at translating basic research results into clinical practice. In the *Cardiac & cardiovascular systems* map (Figure 1), for instance, two bridges are visible, one in the upper part of the map and one in the lower part. In the upper part, the topic of atherosclerosis can be found, starting in the upper-right part of the map with basic research on vascular damage, continuing in the middle part with research on cholesterol and cholesterol lowering drugs, and extending in the upper-left part with interventional therapies such as coronary bypass surgery and percutaneous interventions (PCI) and its modifications (BMS and DES). In the lower part of the map, the topic of arrhythmias can be identified. It starts in the lower-right part of the map with basic research on electrophysiological phenomena, it continues in the middle part with diagnostic tools, and it ends in the lower-left part with the clinical application of ablation therapy for arrhythmias.

Looking at Figures 1, 2, and 3, a crucial observation is that the distinction between different research areas is visible not only in the structure of the maps but also in the colors of the terms. In general, in the right part of each map, in which the more basic and diagnostic research areas are located, there are many yellow, orange, and red terms, which clearly indicates an above-average citation impact. (As indicated by the color bar in the lower right in Figures 1, 2, and 3, yellow and orange correspond with a citation impact that is, respectively, about 25% and about 50% above the average of the field. Red corresponds with a citation impact that is 100% or more above average.) On the other hand, in the left part of each map, research areas can be found with mainly blue and green terms, implying a below-average citation impact. This pattern is most strongly visible in the *Clinical neurology* map (Figure 2) and can also be observed in the *Surgery* map (Figure 3). In the *Cardiac & cardiovascular systems* map (Figure 1), a clear distinction between high- and low-impact research areas is visible as well, but it coincides only partially with the left-right distinction. We further note that within an area in a map terms are usually colored in a quite consistent way. In other words, terms tend to be surrounded mainly by other terms with a similar color. This is an important indication of the robustness of the maps.

The general picture emerging from Figures 1, 2, and 3, and supported by term maps for other medical fields provided online, is that within medical fields there is often a considerable heterogeneity in citation impact, with some research areas on average receiving two or even three times more citations per publication than other research areas. In general, low-impact research areas tend to focus on clinical research, in particular on surgical interventions. Research areas that are more oriented on basic and diagnostic research usually have an above average citation impact.

## 4. Discussion and conclusion

The citation impact of a publication can be influenced by many factors. In the medical sciences, previous studies have for instance analyzed the effect of study design (e.g., case report, randomized controlled trial, or meta-analysis; Patsopoulos, Analatos, & Ioannidis, 2005), article type (i.e., brief report or full-size article; Mavros, Bardakas, Rafailidis, Sardi, Demetriou, & Falagas, 2013), and article length (Falagas, Zarkali, Karageorgopoulos, Bardakas, & Mavros, 2013). In this paper, the effect of differences in citation practices between medical research areas has been investigated.

Different fields of science have different citation practices. In some fields, publications have much longer reference lists than in others. Also, in some fields researchers mainly refer to recent work, while in other fields it is more common to cite older work. Because of such differences between fields, publications in one field may on average receive many more citations than publications in another field.[3] Popular bibliometric indicators, such as the *h*-index and the impact factor, do not correct for this. The use of these indicators to make comparisons between fields may therefore easily lead to invalid conclusions.[4]

The results obtained using the visualization methodology introduced in this paper go one step further and show that even within a single field of science there can be large differences in citation practices. Similar findings have been reported in earlier studies (Neuhaus & Daniel, 2009; Smolinsky & Lercher, 2012; Van Leeuwen & Calero Medina, 2012), but based on smaller analyses and not within the medical domain. The present results suggest that in medical fields low-impact research areas tend to be clinically oriented, focusing mostly on surgical interventions. Basic and diagnostic research areas usually have a citation impact above the field average, although not all high-impact research areas need to have a basic focus. The coloring of the term maps indicates that two- or even threefold impact differences between research areas within a single medical field are not uncommon.

Although differences in citation impact between basic and clinical research have been mentioned in earlier studies (e.g., Seglen, 1997), only a limited amount of empirical evidence of such differences has been collected. We are aware of only a few earlier studies in which differences in citation impact between basic and clinical research have been analyzed (Lewison & Dawson, 1998; Lewison & Devey, 1999; Lewison & Paraje, 2004; Opthof, 2011). These studies are based on much smaller amounts of data than the present analysis. Contrary to the present results, Opthof (2011) concludes that clinical research is cited more frequently than basic research. However, the study is limited in scope. It is restricted to a single medical field, and it considers publications from only a small set of journals.[5] In another relatively small study, reported by Lewison and Paraje (2004), no difference in citation impact between basic and clinical research is detected. This study has the limitation of being restricted to publications from only two journals. Two earlier studies (Lewison & Dawson, 1998; Lewison & Devey, 1999) provide some evidence for a citation advantage for basic publications over clinical ones.

---

[3] It could be suggested that differences in citation impact between research areas may also be caused by the size of an area. In a larger research area, there are more researchers and more publications than in a smaller research area, and therefore there are also more citations. However, one should be careful with this argument. In a large research area, there are many publications, each of them giving citations to earlier work, but at the same time there are also many publications that can be cited. Given this balance between citing and citable publications, one may expect that in general the average number of citations of the publications in a research area is not affected by the size of the area.

[4] This is by no means the only objection one may have against these indicators. An important objection against the impact factor could be that the impact of a journal as a whole may not be representative of the impact of individual publications in the journal (Seglen, 1997). An objection against the *h*-index could be that it suffers from inconsistencies in its definition (Waltman & Van Eck, 2012a).

A number of limitations of the methodology of the present study need to be mentioned. First of all, because the visualization methodology does not draw explicit boundaries between research areas, no exact figures can be provided on citation impact differences between, for instance, basic and clinical research. On the other hand, by not drawing explicit boundaries, many arbitrary choices are avoided and more fine-grained analyses can be performed. Another methodological limitation is the ambiguity in the meaning and use of terms. Some terms may for instance be used both in basic and in clinical research. Although a term selection algorithm was employed to filter out the most ambiguous terms, some degree of ambiguity cannot be avoided when working with terms. Other limitations relate to the bibliographic database that was used. The WoS database has a good coverage of the medical literature, but to some extent the analysis might have been affected by gaps in the coverage of the literature. Also, the analysis depends strongly on the field definitions offered by the WoS database.

The results reported in this paper lead to the conclusion that one should be rather careful with citation-based comparisons between medical research areas, even if in a bibliographic database such as WoS the areas are considered to be part of the same field. Field-normalized bibliometric indicators, which are typically used by professional bibliometric centers, correct for differences in citation practices between fields, but at present they fail to correct for within-field differences. The use of bibliometric indicators, either the *h*-index and the impact factor or more sophisticated field-normalized indicators, may therefore lead to an underestimation of the impact of certain types of research compared with others. In particular, the impact of clinical intervention research may be underestimated, while the impact of basic and diagnostic research may be overestimated.

---

[5] Replicating the two analyses reported by Opthof (2011) confirmed their results. The first analysis reported by Opthof is based on six cardiovascular journals, three basic ones and three clinical ones. The difference between the outcomes of this analysis and the analysis reported in the present paper appears to be related to the particular characteristics of the selected journals. The publications in these journals turn out not to be fully representative for basic and clinical publications in all cardiovascular journals. The second analysis reported by Opthof is based on the distinction between basic and clinical publications within a single cardiovascular journal (*Circulation*). In this case, the difference with the outcomes of the analysis reported in the present paper seems to indicate that the selected journal differs from the cardiovascular field as a whole in terms of the characteristics of its basic and clinical publications.

There is an urgent need for more accurately normalized bibliometric indicators. These indicators should correct not only for differences in citation practices between fields of science, but also for differences between research areas within the same field. Research areas could for instance be defined algorithmically based on citation patterns (e.g., Klavans & Boyack, 2010; Waltman & Van Eck, 2012b). Alternatively, a normalization could be performed at the side of the citing publications by giving a lower weight to citations from publications with long reference lists and a higher weight to citations from publications that cite only a few references. A number of steps towards such citing-side normalization procedures have already been taken (e.g., Glänzel, Schubert, Thijs, & Debackere, 2011; Leydesdorff & Opthof, 2010; Moed, 2010; Rafols, Leydesdorff, O'Hare, Nightingale, & Stirling, 2012; Waltman & Van Eck, in press; Waltman, Van Eck, Van Leeuwen, & Visser, 2013; Zitt & Small, 2008), but more research in this direction is needed. Using the presently available bibliometric indicators, one should be aware of biases caused by differences in citation practices between areas of medical research, especially between basic and clinical areas.

## Acknowledgment

We would like to thank Cathelijn Waaijer for helpful suggestions in the interpretation of the term maps.

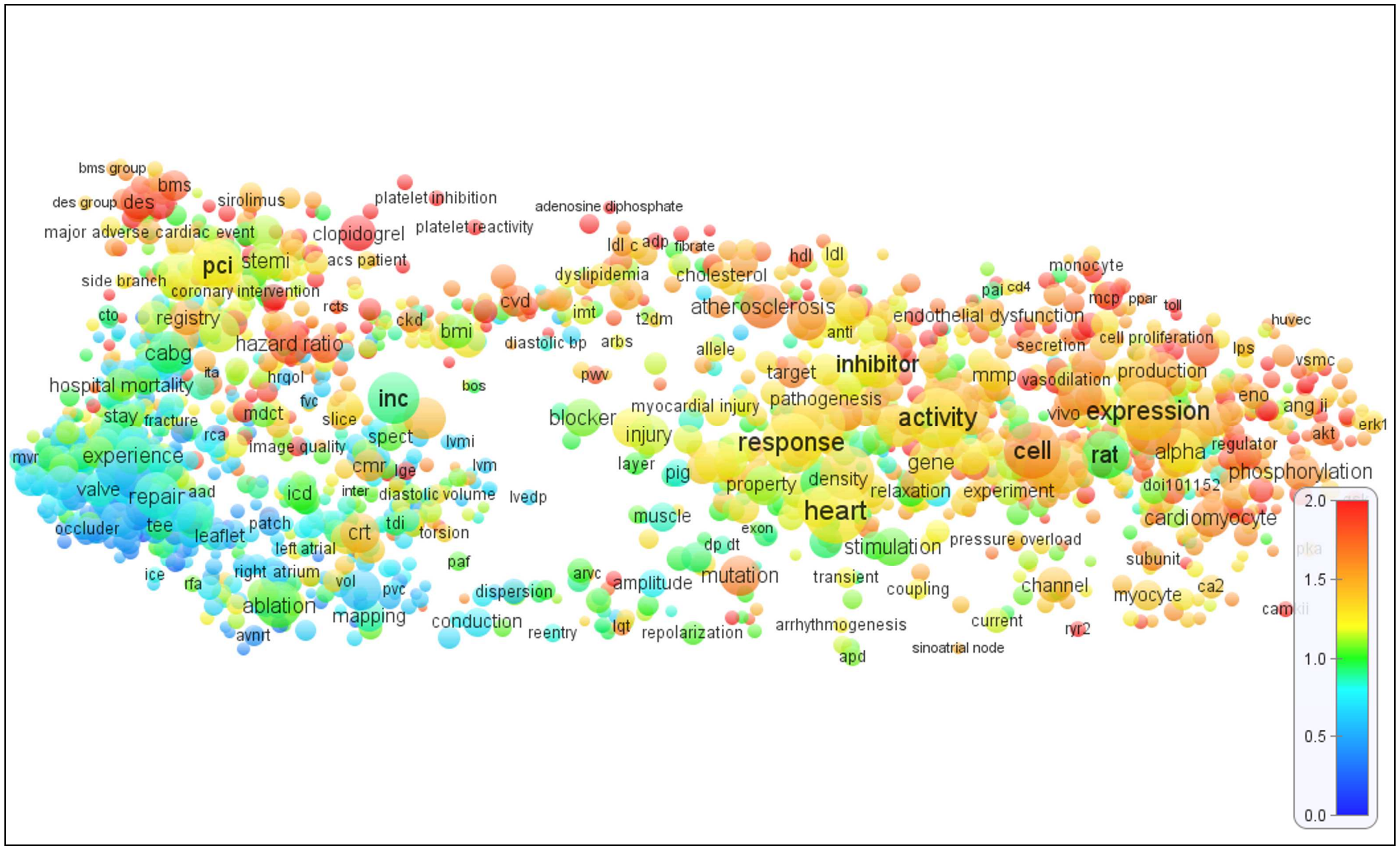

Figure 1. Term map of the *Cardiac & cardiovascular systems* field. The map shows 2000 terms extracted from titles and abstracts of publications in the WoS field *Cardiac & cardiovascular systems*. In general, the closer two terms are located to each other, the stronger their relation. The size and the color of a term indicate, respectively, the number of publications in which the term occurs and the average citation impact of these publications (where blue represents a low citation impact, green a normal citation impact, and red a high citation impact). Each term occurs in at least 70 publications.

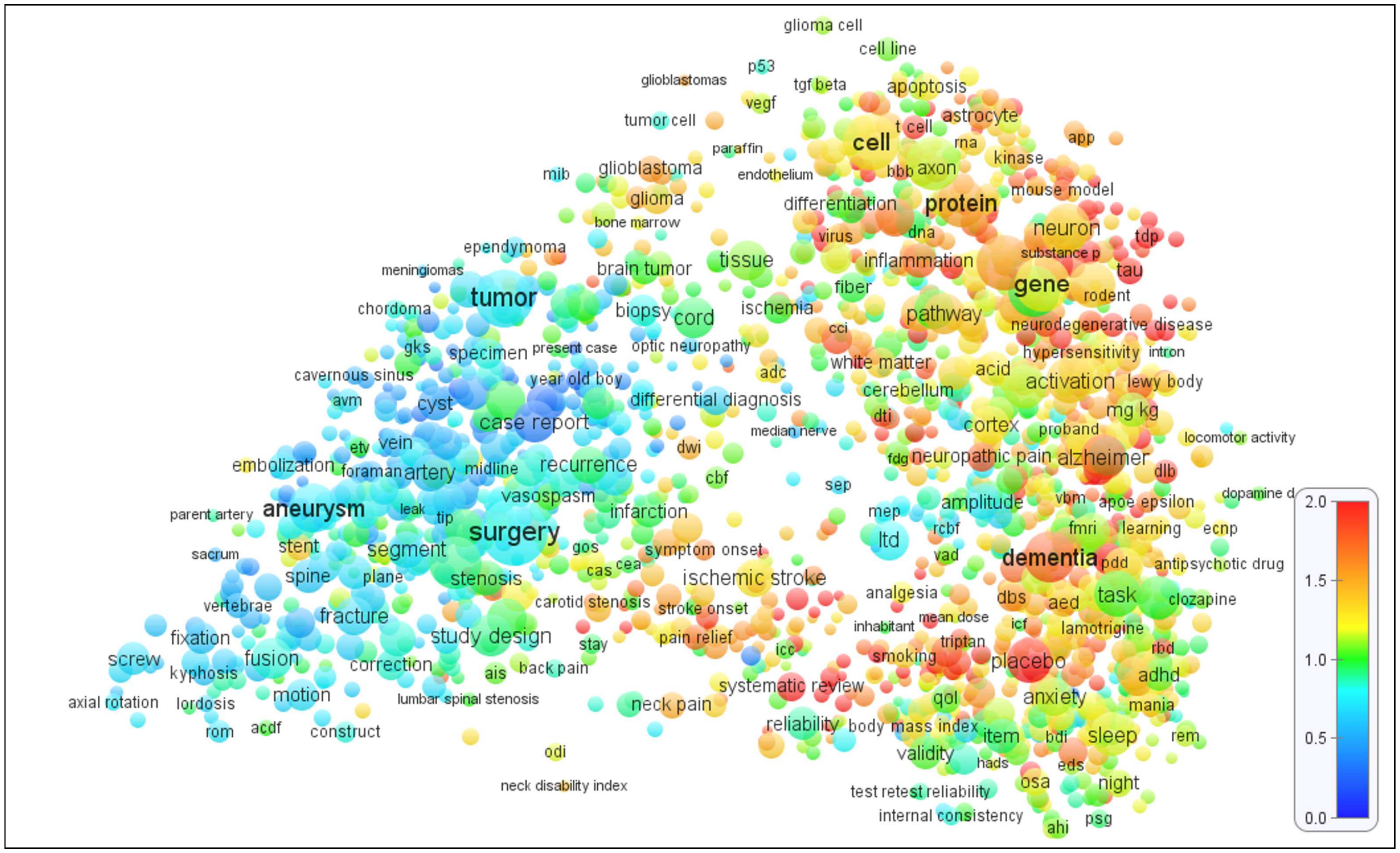

Figure 2. Term map of the *Clinical neurology* field. The map shows 2000 terms extracted from titles and abstracts of publications in the WoS field *Clinical neurology*. In general, the closer two terms are located to each other, the stronger their relation. The size and the color of a term indicate, respectively, the number of publications in which the term occurs and the average citation impact of these publications (where blue represents a low citation impact, green a normal citation impact, and red a high citation impact). Each term occurs in at least 100 publications.

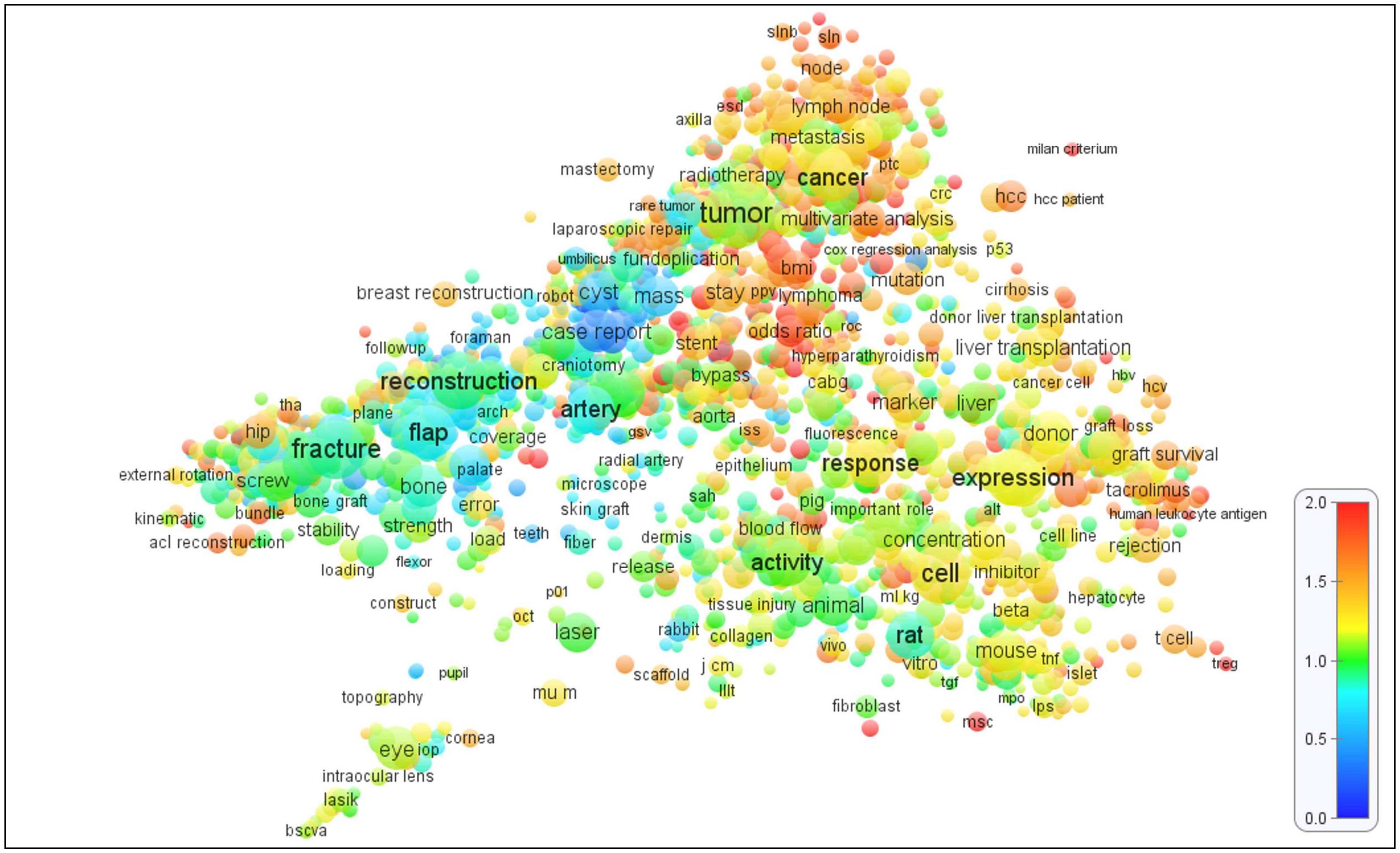

Figure 3. Term map of the *Surgery* field. The map shows 2000 terms extracted from titles and abstracts of publications in the WoS field *Surgery*. In general, the closer two terms are located to each other, the stronger their relation. The size and the color of a term indicate, respectively, the number of publications in which the term occurs and the average citation impact of these publications (where blue represents a low citation impact, green a normal citation impact, and red a high citation impact). Each term occurs in at least 135 publications.